\documentclass[a4paper]{mn2e}
\usepackage{epsfig}
\usepackage[
pdfauthor={M.E. Beer, A.R. King and J.E. Pringle},
pdftitle={The planet in M4: implications for planet formation in globular
  clusters},
pdfsubject={We propose that the planet around the pulsar PSR B1620-26
  formed through the interaction of a passing star with a circumbinary
  disc during the common-envelope phase of the inner binary's
  evolution.},
pdfkeywords={planets and satellites: formation - planetary systems:
  protoplanetary discs - pulsars: individual: PSR B1620-26}
]{hyperref}

\date{Accepted 2004 Sep 15. Received 2004 Aug 18; in original form 2004 June 9}

\volume{000}

\pagerange{\pageref{firstpage}--\pageref{lastpage}} \pubyear{2004}

\begin{document}

\label{firstpage}

\title[The planet in M4: implications for planet formation in globular
  clusters]{The planet in M4: implications for planet formation in
  globular clusters} 

\author[M.\,E.~Beer, A.\,R. King and
  J.\,E. Pringle]{M.\,E.~Beer$^1$\thanks{E-mail:
  martin.beer@astro.le.ac.uk}, A.\,R. King$^1$ and J.\,E. Pringle$^2$
\\
1. Theoretical Astrophysics Group, University of Leicester,
Leicester, LE1~7RH, UK\\
2. Institute of Astronomy, Madingley Rd, Cambridge CB3~OHA, UK
}

\maketitle

\begin{abstract}
We consider the formation and evolution of the planetary system
PSR~B1620--26 in the globular cluster M4. We propose that as M4 is a
very-low metallicity environment the standard model of planet
formation around main-sequence stars through the accretion of gas onto
metallic rocky cores should not be 
applied. Consequently the previously suggested methods for
formation are unlikely. We propose that the planet formed through the
interaction of a passing star with a circumbinary disc during
the common-envelope phase of the inner binary's evolution. This
formation route is favoured by dense stellar systems such as globular
clusters.
\end{abstract}


\section{Introduction}
Extrasolar planets appear to be rare or absent around stars of low
metallicity (Gonzalez 1998; Gonzalez, Wallerstein \& Saar 1999; Reid
2002; Santos et al. 2003; Fischer, Valenti \& Marcy 2004). This has an
appealing 
potential interpretation in terms of the need for a protoplanet to
build a metallic core before accreting further (but see Beer et
al. 2004). Convincing evidence for a planet in a low metallicity
system would clearly offer a challenge to this view.

One potential arena for such a challenge is a globular cluster, where
metallicities are far lower than in any star with a known
planet. There is one known planetary system in a globular
cluster. This is a hierarchical triple with a Jupiter mass planet in a
wide non-coplanar orbit with a binary millisecond pulsar
(PSR~B1620--26) as the inner 
binary (Backer 1993; Thorsett et al. 1999). We propose that
this planet formed through a gravitational instability in a
circumbinary disc as a result of an interaction with a passing star. 

In Section~\ref{prevmodel} we summarise the known properties of the
planet orbiting PSR~B1620--26 as well as the proposed
models for how it formed. In Section~\ref{sectform} we suggest how a
circumbinary disc may form as a result of equatorial outflows during
the common-envelope phase of evolution of the inner
binary. Section~\ref{sectdyn} discusses the likelihood of the disc
encountering a passing star and forming planets. Section~\ref{vesc}
argues that the 
neutron star in the inner binary can only have received a small kick
during formation. This small kick keeps the binary in
the cluster, and the planet bound to it. In
Section~\ref{discuss} we discuss this result and its application to
planet formation in general as well as summarising our conclusions.

\section{The system and previous ideas for its origin} \label{prevmodel}
PSR~B1620--26 is a hierarchical triple system in the globular cluster
M4. The orbital solution has been found by Thorsett et al. (1999). The
inner binary consists of an 11 millisecond pulsar with a white 
dwarf companion. The period and eccentricity of this binary are
191.4\,d and 0.025315 respectively. Observations by Sigurdsson et
al. (2003) indicate that the white dwarf is young and undermassive. The
age and mass of the white dwarf are
4.8\,$\times$\,10$^8$\,$\pm$\,1.4\,$\times$\,10$^8$\,yr and
0.34\,$\pm$\,0.04\,M$_{\sun}$ respectively. The mass of the pulsar is
assumed to be 1.35\,M$_{\sun}$. 

The white-dwarf mass and binary period are in good agreement with the
theoretical predictions for the mass-period relation of Savonije
(1987) and Rappaport et al. (1995). These predictions arise from
assuming the white dwarf is the core of a giant whose envelope was
transferred to the pulsar, spinning it up to millisecond
periods. Stellar evolution theory constrains the envelope radius,
which is filling its Roche lobe in this model, and hence the binary
period as a function of core (white-dwarf) mass. Any model for the
origin of the system should not disrupt this relationship
significantly. 

The third member of this triple is a Jupiter-mass planet whose orbit
has a period of order 100\,yr (Thorsett et al. 1999). This planet was
detected through the observed second period derivative of the
pulsar. The published timing has representative solutions for
different ellipticities of the outer orbit. For an eccentricity of
0.20 the period would be 129\,yr and the minimum companion mass would
be 3.5\,M$_{\rm J}$. If the eccentricity were as large as 0.5 the
period would be 389\,yr and the minimum companion mass would be
6.8\,M$_{\rm J}$. Sigurdsson et al. (2003; 2004) argue that the planet
is at the lower end of the allowed ranges for mass and semi-major
axis. This implies a minimum planet mass of $1.7$\,M$_{\rm J}$, a 
period of 68\,yr and an eccentricity of 0.13. 

Phinney (1992) suggested a relationship between
eccentricity and orbital period for binary millisecond pulsars (see
also Phinney \& Kulkarni 1994). Phinney (1992) argues that this
relationship is caused by the non-central (multipole) forces exerted on
the neutron star due to the fluctuating density of convection cells in
the red-giant progenitor of the white dwarf. The longer the period and
hence the larger the size of the red-giant progenitor the greater the induced
eccentricity. This model fits the observed eccentricity orbital
period distribution of binary millisecond pulsars well. From this model we
would expect an eccentricity of order 10$^{-3}$ for the inner binary
significantly smaller than the observed eccentricity of 0.025. Ford et
al. (2000) have proposed that Kozai pumping by the planetary
companion could explain the observed eccentricity. For such a large
eccentricity Kozai (1962) requires that the difference in inclination
between the inner binary orbit and the planetary orbit is at least
40$^{\circ}$. 

Sigurdsson (1992, 1993, 1995 hereafter S93) has proposed a model for the
origin of the current configuration. In this model the planet forms
around a main sequence star outside the core of M4. The planetary
system migrates towards the core of M4 where it encounters a
neutron-star binary. The neutron star captures the star and planet and
ejects 
its original companion. The system is now a hierarchical triple with
an inner neutron-star plus main-sequence binary and an outer
planet. The main sequence star evolves into a red giant 
and transfers mass to the neutron star. This 
spins it up to millisecond periods, eventually leaving a millisecond
pulsar plus white dwarf binary. 
In this model, during the encounter which forms the hierarchical
triple, 
the system has a recoil which displaces the binary from the core of
the cluster. Sigurdsson et al. (2003) state that a binary sinks into the
cluster core on the relaxation timescale (of order a Gyr). This is
correct but does not give the full picture. As (in
the proposed model) the system was initially in the core, and has
undergone no other exchanges, its orbit takes it back into the core
on the crossing timescale (around a Myr) which is much shorter than
both the relaxation timescale and the age of the white dwarf. For this
model to be correct the system must have formed very recently in order
to avoid encounters which unbind the planet (see
section~\ref{sectdyn}). 

Alternatively, Joshi \& Rasio (1997) proposed that a planetary system
containing a main sequence star encounters a preexisting binary
millisecond pulsar. During this encounter the main sequence star is
ejected, although as the authors note, one would expect the the main
sequence star (as the more massive object) to be retained rather than
the planet. This model can successfully explain the observed inner
binary eccentricity without the requirement for Kozai pumping which
requires a high relative inclination of the two orbits.

A fundamental problem with both of these models, however, is that they
require the 
planet to have already formed in the globular cluster which is a low
metallicity environment. M4 has an [Fe/H] of $-$1.20 (Harris 1996) while
the current lowest known metallicity of a main sequence star with a
planetary companion is HD 6434 which has an [Fe/H] of $-0.52 \pm 0.08$
(Santos et al. 2003). Fischer et al. (2004) have hypothesised that
there is a 
metallicity dependence on planet formation. They find that at solar
metallicity between 5--10\% of stars host Doppler-detected planets. As
stellar metallicity drops toward [Fe/H] of $-$0.5, however, the
occurrence of detected planets declines to a few
percent.

\section{The proposed model} \label{sectform}
\begin{figure}
\begin{center}
\epsfig{file=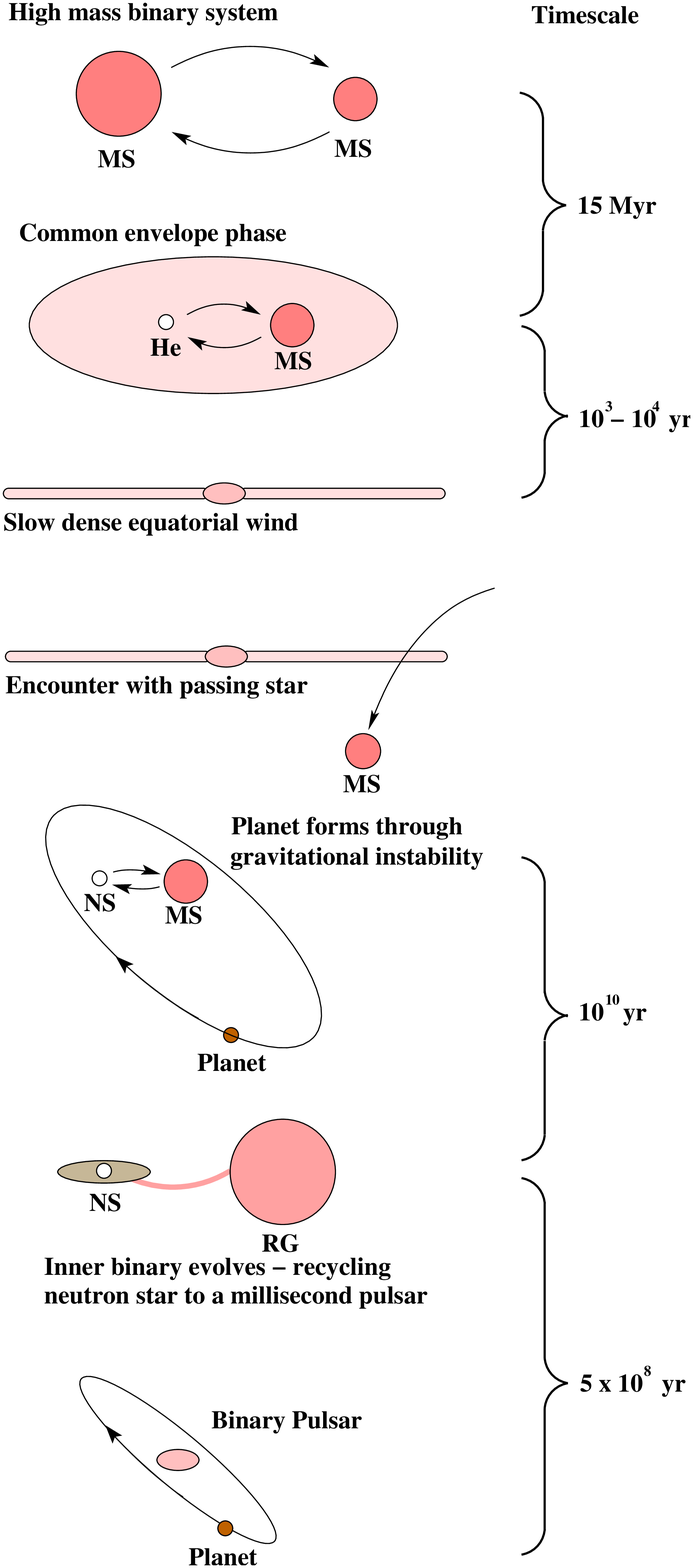,width=7.75cm}
\caption{A figure showing the stages during the formation of the
  hierarchical triple pulsar planetary system and their associated
  timescales. Initially a high mass X-ray binary system consisting of
  e.g. a 15\,M$_{\sun}$ and a 1\,M$_{\sun}$ star evolves on the
  timescale of the 
  most massive component. Once the high-mass component evolves off the
  main sequence mass transfer becomes unstable and a common-envelope
  phase ensues. The common-envelope phase lasts
  $10^3$--$10^4$\,yr. During the common envelope phase a slow, dense
  equatorial wind is formed. A passing star encounters this wind
  and the planet forms
  through a dynamical instability of the disc. The system now consists
  of the progenitor of the binary millisecond pulsar and the
  planet. The system undergoes no further significant encounters on a
  timescale of the age of the cluster $10^{10}$\,yrs. Eventually the
  inner binary evolves spinning up the neutron star to a millisecond
  pulsar and forming the young white dwarf observed today.} \label{figstages}
\end{center}
\end{figure}

Figure~\ref{figstages} shows a schematic view of how the system formed
in the proposed model. During the formation of the inner binary the
progenitor of the 
neutron star underwent a common-envelope phase (see Bhattacharya
\& van den Heuvel 1991 for a discussion of likely evolutionary
scenarios). A typical scenario consists of a binary with a high-mass
component (e.g., 15\,M$_{\sun}$) and a low-mass component
(1\,M$_{\sun}$). At the end of core helium 
burning of the primary its envelope expands until it fills its Roche
lobe and enters a common-envelope phase. During this 
phase the envelope of the neutron star progenitor is
ejected while the white dwarf progenitor spirals in. This leaves the
core of the massive star which is a helium star and the low-mass
companion. The helium star eventually undergoes a supernova leaving
a neutron star possibly in an eccentric orbit (due to the supernova
kick). This orbit circularizes due to tides and if the companion
fills its Roche lobe on the red giant branch then mass
transfer occurs. The red-giant envelope is transferred onto the
neutron star spinning it up to millisecond periods while the
degenerate helium core of the red-giant remains as the white-dwarf
companion. The exact details of the common-envelope phase and 
associated mass loss are not well understood although it is generally
agreed that this phase is short. In the scenario outlined above the
mass loss is of order 10\,M$_{\sun}$ and occurs on a
timescale of 10$^3$--10$^4$\,yr (van den Heuvel 1976).

Possible evidence for the post-common-envelope structure comes from 
observations of planetary nebulae and pre-planetary
nebulae\footnote{Pre-planetary nebulae are also referred to as
  proto-planetary nebulae, but we do not use the phrase here to avoid 
  confusion with the use of proto-planetary nebulae as being
  centrifugally supported 
  discs of accreting matter around stars in which planets may form.}.
In order to explain the shape of many planetary nebulae a slow dense
wind confined to a plane is required prior to a final fast spherically
symmetric wind which interacts with it (see the review of Balick \&
Frank 2002). This would form the observed bipolar outflows observed in
many planetary nebulae (Gieseking,
Becker \& Solf 1985; Miranda \& Solf 1992; L\'opez, Steffen \& Meaburn
1997). This slow, dense wind has been directly observed in some
pre-planetary nebulae through HCN emission e.g., in the `Egg nebula'
(Bieging \& Nguyen-Quang-Rieu 1988) and in the `Westbrook nebula' (S\'anchez
Contreras \& Sahai 2004). In the `Egg nebula' there 
is evidence that this slow wind has significant rotation (see Pringle 1989). 
The `Westbrook nebula' (also known as CRL 618) has been
interferometrically mapped by S\'anchez Contreras \& Sahai
(2004). They inferred  a
dense, equatorial torus expanding at 17.5\,kms$^{-1}$. These 
observations could be of the excited part of a much slower (or even
non-moving) torus. This torus extends out to 5.5\,arcsec which at
their adopted distance of 900\,pc corresponds to an outer radius of
3300\,AU. They estimate a torus mass of 0.25\,M$_{\sun}$ (260\,M$_{\rm
  J}$). 

The
most plausible methods for the formation of this dense, confined gas
are either the expulsion of the wind of a red giant by a binary 
companion in the orbital plane (Morris 1981) or the emergence from
a common-envelope phase (Soker \& Livio 1991; Reyes-Ruiz \& L\'opez
1999). There is a strong resemblance between these winds and
proto-planetary discs (Pringle 1989; Kastner \& Weintraub 1995). There
is no reason why a dense, slow, equatorial wind 
in the progenitor of a binary millisecond pulsar would not be produced 
during/after the common-envelope phase. Two- and three-dimensional
simulations of common-envelope evolution show that the envelope mass
is lost preferentially in directions near the orbital plane (Yorke,
Bodenheimer \& Taam 1995; Sandquist et al. 1998). 

Mastrodemos \& Morris (1998; 1999) performed three-dimensional
hydrodynamical modelling of gas flows in bipolar nebulae including the
formation of accretion discs. We use the term accretion discs loosely
here as the matter in them is not centrifugally supported. In fact the
discs contain significant rotation but are not strongly unbound. The
accretion discs in their simulations  
extended out to between 100 and 1000\,AU and had terminal velocities in
the range 10--27\,kms$^{-1}$. If the general wind loss rate in these
systems is 10$^{-4}$\,M$_{\sun}$yr$^{-1}$ (a typical rate for
planetary nebulae) then a 10\,kms$^{-1}$ wind would have 5\,M$_{\rm
  J}$ within 100\,AU. Simulations by Sandquist et al. (1998)
found similar results with a disc extending out to 100 AU
and a mass $>$\,1\,M$_{\sun}$ although not all the mass in their
simulations was unbound. Such discs have enough
mass at large enough radii to form the planet around
PSR~B1620--26. This planet has a mass of a few Jupiter masses
and a periastron distance of $\sim$30\,AU (Thorsett et al. 1999).

Planet formation in a post-common-envelope disc could proceed by the
standard model for the formation of the solar system giant planets
(Wetherill 1980; Mizuno 1980; Stevenson 1982) if the disc is metal
rich. This standard model assumes that planets form initially through
the agglomeration of dust into grains, pebbles, rocks and thence
planetesimals which form the planetary cores. This
formation route, however, would not provide the large relative
inclination of the two orbits necessary for the Kozai mechanism to
produce the high eccentricity of the inner binary. 
M4 is a low-metallicity cluster but the environment around the
primary which loses its envelope during the asymptotic-giant-branch
primary could be dust rich (Livio \& Pringle 2003). If the planet is
to form bound from the initially unbound matter in this disc/wind
phase an interaction with a passing star is required (see below). 

Gravitational instability in the disc (Boss 2001; Rice et
al. 2003; Mayer et al. 2004) provides another possible
formation mechanism. This instability requires a change in disc 
properties on a timescale comparable to or less than the dynamical timescale
of the disc. Interaction with a passing star would provide this and is
likely in a globular cluster environment. Any encounter would probably
be non-coplanar and so we expect the resulting planets to be
non-coplanar as well. Detailed numerical simulations are required,
however, to verify if this formation route could indeed produce the
planet around PSR~B1620--26. Gravitational instabilities typically
generate many planetary mass objects (Rice et al. 2003). Some of these
may escape from the system as a result of the encounter while others
in the case of PSR~B1620--26 may not yet be detectable.

Hall, Clarke \& Pringle (1996) have investigated the response of a
circumstellar accretion disc to the fly-by of a perturbing star on a
parabolic orbit. They considered a perturbing star of equal mass to
the central star and a disc of non-interacting
particles i.e. they did not include hydrodynamical forces in the
disc. Their analysis investigated which regions of the disc remain
bound after an encounter in addition to whether energy and angular
momentum is gained/lost from the perturbing star. They found that for
prograde fly-bys matter outside of periastron is lost from the system
(i.e. made unbound) while matter within 60 per cent of the periastron
distance is not only retained but because of a corotation resonance
loses energy and angular momentum to the binary orbit.\footnote{Here
  binary orbit refers to the orbit of the two stars which is initially
  parabolic i.e. has an eccentricity of 1.} In retrograde fly-bys
there is no corotation resonance and the binary orbit loses energy to the
disc principally in the sense of matter outside of periastron becoming
unbound from the system. 

\begin{table*}
\begin{center}
\begin{tabular}{l c c}
\hline
Parameter & Value & Reference \\
\hline
Distance & 2.2\,kpc & Harris (1996) \\
Core radius ($r_{\rm c}$) & 0.83$'$ & Trager, Dvorgovski \& King (1993) \\
Half-mass radius ($r_{\rm h}$) & 3.65$'$ & Harris (1996) \\
Tidal radius ($r_{\rm t}$) & 32.49$'$ & Harris (1996) \\
Absolute magnitude ($M_{\rm V}$) & $-$7.20 & Harris (1996) \\
Velocity dispersion ($\sigma_{\rm obs}$) & 3.88\,$\pm0.64$ & Peterson \& Latham (1986) \\
& 4.44\,$\pm0.71$ &  Rastorguev \& Samus (1991) \\
& 3.50\,$\pm0.2$ & Peterson, Rees \& Cudworth (1995) \\
Concentration ($c$) & 1.59 & Trager, Dvorgovski \& King (1993) \\
Metallicity [Fe/H] & $-$1.20 & Harris (1996) \\
\hline
\end{tabular}
\caption{A table listing the properties of the globular
  cluster M4.} \label{tablem4}
\end{center}
\end{table*}

Boffin et al. (1998) have simulated star-disc
encounters using a smoothed particle hydrodynamical approach. They
find that the formation of spiral arms and fragmentation occur in the
circumstellar disc. The spiral arms have density
contrasts of greater than two orders of magnitude compared to the
initial disc and it is in these that fragmentation occurs. Pfalzner
(2003) has also simulated star-disc encounters and finds formation of
spiral arms in the disc. Both of these simulations had a limited
number of particles (11,300 and 50,000 particles respectively) and
higher resolution simulations of star-disc encounters need to
be carried out to investigate the possible fragmentation of the disc
due to an encounter further.

In non-coplanar encounters the passing star imparts a perturbation to
the accretion disc in its vertical direction (Heller 1993). This tilts
the accretion disc compared to its unperturbed orientation. If the
disc were to fragment due to this encounter then the planets formed
would have inclined orbits relative to the initial disc.

In the model it is easy to envisage a prograde encounter with a
periastron greater than the the disc radius in which the planet forms
(of order 100\,AU) shocking material inside of this radius and
extracting energy and angular momentum from this material making it
bound. It is in this region that the observed planet can
form. Detailed simulations are required to verify this scenario but
those already performed demonstrate the overdensities which may be
induced in addition to the extraction of energy from the disc matter.
Indeed the initial binary is so massive that the primary evolves
before the cluster is fully formed. The encounter which forms the
planet could occur during this phase.

\section{Dynamical considerations} \label{sectdyn}
\begin{figure}
\begin{center}
\epsfig{file=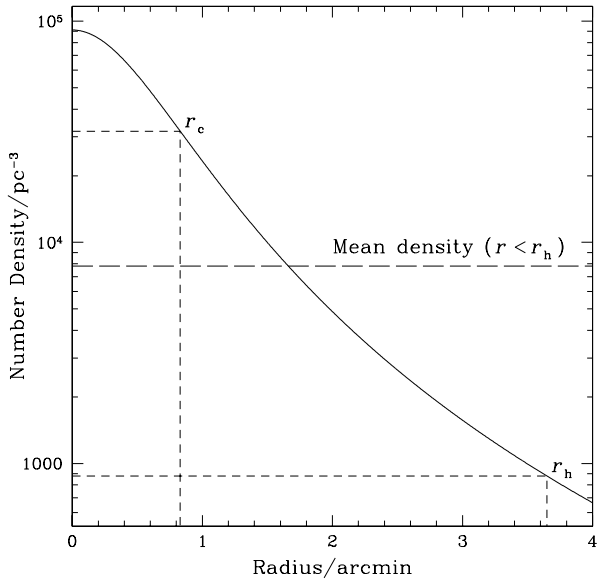,width=8.25cm}
\caption{A figure showing a King model for the density profile of the
  globular cluster M4. The long-dashed line shows the mean density
  inside the half-mass radius and the short-dashed lines the positions
  and densities at the core radius ($r_{\rm c}$) and half-mass radius
  ($r_{\rm h}$).}
  \label{figkingprofile}
\end{center}
\end{figure}
In this section we discuss the possible dynamical history of
PSR~B1620--26. The aim of this section is to show that the system
could be outside the core of M4 and that it would {\it not} have sunk
into the core of the core within a Hubble time. We start by
considering the possible position of the system in the cluster with
respect to its observed position which is within the core. We then
discuss the relaxation time for an object in M4 to sink into the core
using a King profile for the radial number density. We also
discuss the encounter timescale for the system and the effects
encounters have on the planetary orbit which may initially have been
tighter than currently observed.

\subsection{Relaxation into cluster core}
The observed offset of PSR~B1620--26 from the core of M4 (0.767 arcmin)
is less than 
the observed core radius (0.83 arcmin) and considerably less than the
half mass radius (3.65 arcmin). This could simply be a projection
effect however, with the system actually far outside the core. This
is required in both the model described here and that in S93. This would
not be unprecedented for globular cluster pulsars e.g. pulsars A and C
in NGC 6752, PSR J1748--2444 in Terzan 5 and PSR B1718--19 in NGC 6342
all lie outside the half mass radius (see Freire 2004 for a list of
the observed properties of globular cluster pulsars). The chance of
the system having a projected distance ($r_{\rm app}$) less than the
core radius ($r_{\rm c}$) when
it is actually at the half-mass radius ($r_{\rm h}$) is given by
\begin{equation}
f(r_{\rm app}<r_{\rm c}) = 1 - \cos \left[ \arcsin(r_{\rm c}/{r_{\rm
    h}}) \right] ~,\
\end{equation}
which for M4 is 8 per cent.

The relaxation time for an object to enter the core of the 
globular cluster is inversely proportional to the stellar density (see
Djorgovski 1993). The half-mass relaxation time for M4 is of order
1\,Gyr (Harris 1996). However, this assumes a stellar density which is
equivalent to the mean stellar density inside the half-mass
radius. 
King (1962) finds the density profile of a globular cluster to be
\begin{equation}
\rho(r) \propto \frac{1}{z^2} \left [ \frac {\arccos z}{z} -
  \sqrt{1-z^2} \right ] ~,\
\end{equation}
where
\begin{equation}
z = \sqrt{\frac{1+(r/r_{\rm c})^2}{1+(r_{\rm t}/r_{\rm c})^2}} ~,\
\end{equation}
and $r_{\rm t}$ is the tidal radius of the
globular cluster. Figure~\ref{figkingprofile} shows the
density profile of the globular cluster M4 out to 4 arcmin. Also shown
on figure~\ref{figkingprofile} are the mean number density inside of
the half-mass radius (calculated below) and the densities at the core
and half mass radius respectively. 
Outside the half-mass radius the stellar density is much
less than the mean density inside the half-mass radius (at least an
order of magnitude) and so the
relaxation time is much greater. Indeed a system outside the
half-mass radius would not necessarily enter inside the half-mass
radius within a Hubble time. 
We conclude that if the system was formed outside the half-mass radius
of the cluster then the system would still be outside this radius and
would not have relaxed into the cluster core.

\subsection{Stellar encounters}
During the initial formation stages of a globular cluster there occurs
a phase of residual gas expulsion which occurs on a timescale of
$10^7$\,yrs (Meylan \& Heggie 1997). Irregular mass loss may cause a
phase of dynamical 
mixing - the violent relaxation stage (Lynden-Bell 1962; 1967). This
violent relaxation occurs on a timescale of a few crossing times for
the globular cluster. The crossing time of the cluster is of 
order $10^6$\,yrs so the violent relaxation timescale is comparable to
the 15\,Myr timescale on which the primary evolves and the planet
formed in the model described here. Therefore the encounter required
to form the planet in the model may have occured during this violent 
relaxation stage (an encounter within $\sim$100\,AU is all that is
required). Indeed given the number of binary millisecond pulsars in
globular clusters it is possible that more have undergone a similar
formation route and have planetary mass companions in wide
non-coplanar orbits. In low density clusters like M4 these planets may
survive without being disrupted by stellar encounters (Davies \&
Sigurdsson 2001). 

We now consider the timescale for the system (including the planet) to
encounter another star and what the outcome of this encounter will
be. Davies \& Sigurdsson (2001) give the timescale for an encounter
between a binary and a single star as
\begin{equation}
\tau_{\rm enc} = \frac{1}{n\sigma v} ~,\
\end{equation}
where $n$ is the number density of stars, $v$ is the velocity of the
binary (which we take to be the velocity dispersion of M4) and
$\sigma$ is the cross-section for two stars having a relative velocity
at infinity of $v_{\infty}$ to pass within a distance $b_{\rm min}$ of
one another
\begin{equation}
\sigma = \pi b_{\rm min}^2 \left ( 1 + \frac{2\,G\,M} {b_{\rm
      min}\, v_{\infty}} \right ) ~,\
\end{equation}
where $M$ is the total mass of all the components and the second term
on the right is due to gravitational focusing and dominates in the
low dispersion environment of a globular cluster. We assume that
$v_{\infty}$ is similar to the velocity dispersion (see
Section~\ref{vesc}). The encounter timescale may now be written
\begin{equation}
\tau_{\rm enc} = 7 \times 10^{8} \frac {10^5\,{\rm pc}^{-3}}{n} \frac
     {{\rm AU}} {b_{\rm min}} \frac{{\rm M}_{\sun}}{M}
    \frac{v_{\infty}}{10\,{\rm kms}^{-1}} \,{\rm yr}~.\
\label{eqtenc}
\end{equation}
We are interested in encounters within 100\,AU. These can affect the
planet without perturbing the inner binary. 
Following Djorgovski (1993) and Harris (1996) the mass of the cluster
($M_{\rm cl}$) is given by
\begin{equation}
\log \,(M_{\rm cl}/2) = 0.4\,(4.79-M_{\rm V}) ~\,
\end{equation}
where $M_{\rm V}$ is the cluster absolute magnitude and a mass to 
light ratio of 2 has been assumed. Table~\ref{tablem4} contains a
list of the properties of the cluster M4. The absolute magnitude
is $-$7.2 which
corresponds to a cluster mass of $1.25\times 10^5$\,M$_{\sun}$. The
mean number density ($\overline{N}$) inside the half-mass radius is given by 
\begin{equation}
\overline{N} = \frac{3\,M_{\rm cl}}{4\pi\,\overline{m} \,r_{\rm h}^3} ~\,
\end{equation}
where $\overline{m}$ is the mean stellar mass (taken to be
1/3\,M$_{\sun}$ following Harris 1996)
and $r_{\rm h}$ is the half mass radius which is 2.34\,pc using the 
distance to M4 of 2.2\,kpc. This gives a mean number density inside the
half-mass radius of 8$\times$10$^3$\,pc$^{-3}$. 

If the number density in the region of M4 where the system currently is
of order 10$^3$\,pc$^{-3}$ then equation~(\ref{eqtenc}) gives the
encounter timescale as 150\,Myr. The exact distance of the system from
the core of M4 is unknown so the timescale may be even larger than
this. 
Indeed the initial orbit of the planet once formed may have been
tighter than that currently observed (because of the softening effects of
encounters - see below). Consequently the encounter timescale (for an
encounter of closest approach comparable to the planetary orbit) may
have been larger as well. If the system were at the core radius then
the timescale for an encounter would be much less (4.5\,Myr) as the
number density is higher. This demonstrates that the planet is far
more likely to survive if it is outside of the core of M4.

During encounters there is a tendency to equipartition energy between
the total kinetic energy of the objects involved and the internal
energy of the system (Heggie \& Hut 2003). Hard systems have greater
internal energy so encounters result in a tighter system and greater
kinetic energy. Soft systems have greater kinetic energy which
results in a widening of the system. So during encounters, hard systems
get harder and soft systems get softer. The boundary between these two
regimes is known as the hard/soft boundary ($R_{\rm hs}$) and is given
by (Bonnell et al. 2001)
\begin{equation}
R_{\rm hs} = \frac {G M_1 M_2 (M_1 + M_2 +M_3)} {M_3 (M_1 + M_2)
  v_{\rm enc}^2} ~,\
\end{equation}
where $M_1$, $M_2$ and $M_3$ are the masses of the inner binary,
planet and encountered star respectively and $v_{\rm enc}$ is the
encounter velocity which may be taken to be the same as the velocity
dispersion for the cluster. This gives the hard/soft boundary for the
planet as a separation of under 1\,AU. The planet in this system is
consequently always softly bound and encounters/flybys on average
lead to it becoming less strongly bound. The outcomes of stellar
encounters has been deeply investigated in the case of equal mass
components (Hut 1984; Heggie \& Hut 1993), but only a few
investigations have been carried out for planetary systems (Bonnell et
al. 2001; Davies \& Sigurdsson 2001; Woolfson 2004).

In their calculations Bonnell
et al. (2001) assumed that if a planet was soft it became unbound
during an encounter. This is not necessarily the case as Woolfson
(2004) has shown when he considered the survivability of planets in
wide orbits. Woolfson (2004) considered an open cluster
($v_{\rm enc} \sim$ 1\,kms$^{-1}$) and integrated for 10$^7$\,yrs. For
planets with $a$ of order 100\,AU he found a large survival
probability even for encounters with periastron distances of a similar
size. This is because the outcome of an encounter depends on the
position of the planet in its orbit and the relative orientation of
the planetary orbit and the path of the encountered star. Woolfson
(2004) found that encounters tendered to soften rather than harden
orbits as expected but with little change in
energy. Consequently it is possible for the planet to undergo a number
of encounters without unbinding it from the system. More
numerical simulations are required, however, to verify this scenario
before we can authoritatively say what the timescale for disruption of
the hierarchical triple is.

In the scenario described above after the planet has formed the inner
binary evolves to form the 
binary millisecond pulsar observed today. As this binary is wide
(191\,d) the accretion onto the neutron star to spin it up to
millisecond periods is transient i.e. long quiescent
periods during which no accretion occurs and short outburst periods
during which accretion occurs but the accretion rate is
super-Eddington (see Taam, King \& Ritter 2000). During these
outbursts of super-Eddington accretion most of the transferred matter
is expelled from the system while a small amount is accreted by the
neutron star spinning it up to millisecond periods. The ejection of
matter from the inner binary results in the planet moving further
out. The change in semi-major axis of the planet due to mass lost is
given by
\begin{equation}
a = a_{\rm i} \frac{M_{\rm i}}{M} ~,\
\end{equation}
where $a$ and $M$ are the semi-major axis of the planet and total mass of
the triple and the subscript i refers to their respective values prior
to mass transfer and spin-up of the neutron star. If half a solar mass
was ejected from the system to give the pulsar (assumed mass
1.35\,M$_{\sun}$) and the 0.34\,M$_{\sun}$ white dwarf then the
semi-major axis increased by 30\,\%. Although this only
represents a small factor in encounter probability it means that the
encounter timescale was longer until recently (as the white dwarf is
young) which may help explain its survival.

\subsection{How radial is the orbit?}
In the model described above the system formed outside the
half-mass radius of the cluster and has an orbit which does not take
it into the core i.e. the orbit does not have a strong radial
component. This is a prediction of the model and may be testable in
future.

In the model described in S93 the system has been ejected
from the core of the cluster due to an exchange encounter. From here
the planet describes an orbit which takes it out of the core and
back in again. The period of this orbit is similar to a crossing
time (of order a Myr). It could be argued that although the system's
orbit takes it back into 
the core on a Myr timescale the system does not spend much time in the
core as it passes through quickly. Consequently it is both more likely
to be found outside the core and less likely to have undergone an
encounter which unbinds the planet from its orbit. 

If the system had a radial orbit, although it is more likely to be
found outside the core, we should still expect it to have undergone a
number of encounters in the core by now. This can be seen by the following
simple argument. Assuming the orbit is entirely radial, in a 
medium with a density equal to the mean density inside the half-mass
radius, and that it orbits between the core and the half-mass radius,
the equation of motion is
\begin{equation}
\frac{d^2 r}{d t^2} = -\frac{G M_{\rm cl} m }{2\,r_{\rm h}^3}\, r ~,\ 
\end{equation}
and the system is a simple harmonic oscillator. It is simple to
show that the system spends a fraction $\arcsin(r_{\rm
  c}/r_{\rm h})$ equal to 13 per cent of its orbit inside the core
radius. Using the age of the white dwarf as a lower limit to the
interaction timescale this corresponds to 62.4 Myr. As we show above
it is probable that the system would have 
undergone a number of encounters in the core of the cluster within
this timescale. 

The main motivation for this paper, however, remains how the planet
could have formed in a low metallicity environment and not whether the
model proposed by S93 would have been disrupted in the core.

\section{A low-kick velocity for the neutron star} \label{vesc}
Before accepting the picture sketched above, we should check that the
system would not be unbound by the natal supernova kick felt by the
neutron star. 
Gnedin et al. (2002) find that the relation between escape velocity
($v_{\rm esc}$), velocity dispersion ($\sigma_{\rm obs}$) and concentration
($c$) of a globular cluster is given by
\begin{equation}
\frac{v_{\rm esc}}{\sigma_{\rm obs}} = 3.7 + 0.9 (c -1.4) ~.
\end{equation}
The data shown in Table~\ref{tablem4} then give the mean velocity
dispersion of 4\,kms$^{-1}$ and an escape
velocity of 15.25\,kms$^{-1}$ 
from the core of M4. Any kick the neutron
star received must have been less than this or the binary would have
escaped from the cluster. Consequently the neutron star kick must have
been small. Pfahl et al. (2002) have argued that there is a population
of neutron-star binaries which receive low kicks. The neutron star in
PSR~B1620--26 may either have received one of these low kicks or could
be at the low-velocity end of the high-kick velocity distribution.

The planet is not strongly bound at formation, 
so we might expect that a neutron-star kick would 
unbind it. The neutron-star
kick, however, may be towards the planet rather than away from
it, making the planet more strongly
bound. Consequently, the presence of a planet in a wide orbit around
the inner binary does not rule out the formation of the planet prior
to the neutron-star kick.

\section{Discussion} \label{discuss}
We have suggested that the formation of the planet in the globular cluster
M4 is possible through a gravitational instability in a circumbinary
disc caused by an encounter with a passing star. This formation
process is metallicity-independent and demonstrates that it is
possible to form planets in globular cluster environments. We stress,
however, that numerical simulations are required to investigate the
possible outcomes of such an encounter. As the formation mechanism
requires an encounter with a passing star, which is unlikely outside a
cluster, we predict that similar objects are not in the field. 

There is one other planetary system known around a
pulsar. PSR~B1257+12 has three Earth mass planetary companions
(Wolzczan \& Frail 1992). These are thought to have formed through
the disruption of the very-low-mass companion of a binary millisecond
pulsar through a 
dynamical instability (Stevens, Rees \& Podsiadlowski 1992; King et
al. 2004). Once the low-mass companion is disrupted an accretion disc
forms around the pulsar from which the planets are formed. Since the
companion contained nuclear-processed material of high metallicity,
Earth-like planet formation is possible. This
process also occurs in a globular cluster although it cannot
produce planets with as large an orbit as that in PSR~B1620--26 (which
has a periastron distance of 30\,AU), or as heavy a mass (all the planets
orbiting PSR~B1257+12 have Earth masses).

To conclude, the globular cluster M4 is a very-low metallicity
environment and so 
the standard model of planet formation should not be applied to the
observed planetary system in it. Consequently the previously suggested
methods for formation are unlikely. The Jupiter mass planet observed
in M4 is a member of a hierarchical triple. When 
the inner binary underwent a common-envelope phase a circumbinary disc
formed as a slow, dense, equatorial wind. This disc extended  
beyond 100\,AU and we propose it underwent an encounter
with a passing star. This 
encounter caused a gravitational instability in the disc and
consequent planet formation. This planet was formed in a wide, 
eccentric, non-coplanar orbit around the inner binary. No other
encounters/exchanges are required to explain the formation of this
system.

\section*{Acknowledgements} 
We thank the anonymous referee for comments which have improved this
paper. 
Theoretical astrophysics research at Leicester is supported
by a PPARC rolling grant. ARK gratefully acknowledges a Royal Society
Wolfson Research Merit Award. MEB acknowledges the support of a UKAFF
fellowship.

\label{lastpage}

\bsp

\end{document}